\documentclass[runningheads]{llncs}
\usepackage{graphicx}
\usepackage{lambda}
\usepackage{bcprules}
\usepackage{amsmath}
\usepackage{amssymb}
\usepackage{listings}
\usepackage{xcolor}
\usepackage{soul}
\usepackage{enumitem}

\newcommand{\cdarkg}[1]{\text{\colorbox{lightgray}{$#1$}}}

\newcommand{\colive}[1]{\textcolor{olive}{#1}}

\newcommand{\ccyan}[1]{\textcolor{cyan}{#1}}

\definecolor{dkgreen}{rgb}{0,0.6,0}

\AtBeginDocument{%
  \setlength\abovedisplayskip{0.2\abovedisplayskip}%
  \setlength\belowdisplayskip{0.2\belowdisplayskip}%
  \setlength\abovedisplayshortskip{0.2\abovedisplayshortskip}%
  \setlength\belowdisplayshortskip{0.2\belowdisplayshortskip}%
  \setlength\floatsep{0.2\floatsep}%
}

\begin{document}
\title{Towards Intrinsic Definitional Interpreters for \\ Hoare Logics}
%
%
\author{Ke Sun\inst{1} \and Di Wang\inst{1} \and Yuyan Bao\inst{2} \and
Meng Wang\inst{3} \and Dan Hao\inst{1}}
\authorrunning{K. Sun et al.}
%

\institute{Key Lab of HCST (PKU), MOE; SCS, Peking University, China
\email{\{sunke@stu.,wangdi95@,haodan@\}pku.edu.cn}
\and
Augusta University, USA\\
\email{yubao@augusta.edu}\and
University of Bristol, UK\\
\email{meng.wang@bristol.ac.uk}
}

\maketitle              
\begin{abstract}
Intrinsic definitional interpreters—definitional interpreters that operate on typing derivations instead of abstract syntax trees—have recently been studied as a promising methodology for defining dynamic semantics of programming languages. A key benefit is that type safety interactively guides and constrains the interpreter’s construction. Analogously to typing relations, Hoare logic is widely used to guarantee program correctness. Can intrinsic definitional interpreters be realized to operate over Hoare-logic derivations?

We explore this question in depth by developing definitional interpreters in Rocq for (i) a basic Hoare logic, and (ii) a realistic logic featuring heaps, dynamic-frame-based local reasoning, well-founded functions, and behavioral subtyping. Central to our approach is a novel technique we call \emph{entry-indexing}, which we use to interpret total-correctness derivations and well-founded functions. Our second development yields, to our knowledge, the first formalization of a dynamic-frame-based Hoare logic with well-founded functions, behavioral subtyping, and total correctness, as well as the first fully mechanized Hoare logic with dynamic frames.
\keywords{Hoare Logic  \and Definitional Interpreter \and Rocq}
\end{abstract}
\section{Introduction}


Definitional interpreters, typically written in functional languages (the \emph{host} languages), have been widely used to specify the dynamic semantics of programming languages (the \emph{object} languages)~\cite{midtgaard2013engineering,owens2016functional,reynolds1972definitional,reynolds1998definitional,yang2019fether}. Unlike small- or big-step semantics, they rely solely on functional constructs and are executable.

Once we have a definitional interpreter for a \textit{strongly typed} object language, we want to prove that it agrees with the language’s typing relation. Suppose we use a dependently typed functional host language, in which typing relations can be encoded as inductive types. The traditional approach to establishing this agreement is to write separate inductive proofs~\cite{amin2017type,ernst2006virtual,siek2013three}, which are typically long and error-prone. However, in recent years, \emph{intrinsic definitional interpreters} (abbreviated as IDIs) have attracted much attention~\cite{bach2017intrinsically,van2022intrinsically,rouvoet2020intrinsically,augustsson1999exercise}. The basic idea behind an IDI is to define interpreters directly over typing derivations. The key benefit of IDIs is that the typing derivations guide the construction of the interpreter interactively, ensuring that the resulting interpreter agrees with the typing relation—and thereby achieves type soundness—by construction.


Like typing relations, Hoare logics are important tools for defining static semantics~\cite{hoare1969axiomatic,almeida2011rigorous,ahrendt2016deductive,nipkow2014concrete}. In dependently typed languages, they are typically encoded as inductive types~\cite{nipkow2014concrete,von2002hoare}. However, to the best of our knowledge, no prior work has explored defining IDIs for Hoare logics; we investigate this possibility in depth in this paper. In addition to adapting the IDI methodology, we set up the following two design goals for our Hoare-logic IDIs.

First, similar to typing rules, we expect that Hoare-logic inference rules can directly \textbf{guide and constrain interpreter construction}. Consider the type signature (in Rocq syntax) shown below: The interpreter operates on partial-correctness derivations ($\{P\}c\{Q\}$). To model non-terminating behaviours with a total function, it takes a parameter \textit{step}. Given a state that satisfies the precondition, it returns a state that satisfies the postcondition, or $None$ if it fails to terminate. \footnote{\textit{asn} is shallowly embedded as $state\rightarrow$ Prop, while $\{v:T|P\ v\}$ is the Rocq notation for a dependent pair: a value of type $T$ and a proof that $P\ v$ holds.} Hence, the signature expresses precisely the soundness theorem for partial-correctness derivations: the interpreter is sound by construction.


\begin{lstlisting}[mathescape]
interP: {P}c{Q}->nat->{v:state|P v}->option {v:state|Q v}.
\end{lstlisting}

Second, we expect that IDIs for total-correctness derivations \textbf{admit totality without step-indexing}. Normally, if we want to define definitional interpreters for object languages with recursive procedures---in a dependently-typed programming language---we would have to use step-indexing to ensure totality~\cite{siek2013three,amin2017type,bach2017intrinsically}. However, step-indexing can be cumbersome in many situations: if we want to execute the interpreter with some concrete input, we must ``guess'' a step that is large enough to make the interpreter terminate; and if we want to do some proof about this interpreter, we must always consider the possibility of $None$ values. Finally, the soundness theorem for total-correctness derivations precludes non-termination, so a faithful interpreter should should dispense with step-indexing. Our major technical contribution is the design of a general technique, dubbed \textit{entry-indexing}, to achieve this. The core of our technique is \textit{entry-indexed total-correctness derivations}. The signature below illustrates this idea. It is similar to that of $interP$, but we interpret entry-indexed total-correctness derivations, written $q;t\triangleright[P]c[Q]$ (explained in Section~\ref{sec:simple}). The parameter $step$ and the \textit{option} in the return type are no longer present. We call IDIs constructed this way as entry-indexed intrinsic definitional interpreters, \textit{E-IDIs}. 
\begin{lstlisting}[mathescape]
interT: q;t$\triangleright$[P]c[Q]->{v:state|P v}->{v:state|Q v}.
\end{lstlisting}


In practice, only some components are provably terminating. We introduce a rule casting total- to partial-correctness derivations, and the partial-correctness interpreter delegates those to the total-correctness interpreter. Thus, it is able to handle \textit{mixed} programs without giving up the benefits of total correctness: only non-terminating components consume steps.

As concrete realizations of these ideas, we present two Hoare logics and develop their partial- and total-correctness IDIs. The first is a very basic logic~\cite{almeida2011rigorous} that will be briefly discussed in Section~\ref{sec:simple} to give an overview of the technique. The second is for a more realistic programming language with heaps, classes, and traits. We use it to exercise the applicability of E-IDI. The main methodology for the second Hoare logic is dynamic frames (DF)~\cite{kassios2006dynamic,kassios2011dynamic}.\footnote{We leave other methodologies (e.g., separation logic) for future work.} To accommodate the more complicated Hoare logic, we substantially extend the entry-indexing technique (introduced in detail in Section~\ref{sec:towards}): (i) it interprets \textit{weakest-precondition proofs} to support DF-style expressions/assertions with well-founded functions; (ii) it is equipped with the notion of \textit{virtual entries}, to account for dynamic dispatch; and (iii) it is extended with \textit{measure-based termination}, suitable for modeling DF-style termination proofs~\cite{leino2009specification,leino2010dafny}. We implement XRL (the Executable Region Logic) using these ideas. 
To our knowledge, it is the first DF-based Hoare logic that supports well-founded functions, dynamic dispatch, and total correctness, and whose soundness is proved w.r.t. a formal semantics.



The two Hoare logics and their IDIs are fully mechanized in Rocq. The paper example and some additional examples, are included in the Rocq development. To our knowledge, this yields the first fully mechanized DF-based Hoare logic. This development is accessible via \url{https://zenodo.org/records/17828628}. 


To summarize, this paper makes the following contributions: 
\begin{itemize}
    \item An entry-indexing technique to construct total definitional interpreters in the presence of mutual recursion and dynamic dispatch, as well as a methodology to interpret Hoare logic derivations mixing total and partial correctness;
    \item An in-depth exploration of definitional interpreters for Hoare logics, resulting in the first rigorous DF-based Hoare logic that supports well-founded functions, dynamic dispatch, and total correctness; and
    \item A full mechanization of the development in Rocq, resulting in the first fully mechanized DF-based Hoare logic.
\end{itemize}

\section{Overview}
In this section, we give an overview in two steps. First, we illustrate the technique with a simple Hoare logic. Then we revisit dynamic frames (DF) and identify the challenges of applying the technique to DF-based Hoare logics.


\subsection{An E-IDI for A Simple Hoare Logic}
\label{sec:simple}
\begin{lstlisting}[mathescape]
Definition state := var -> val.
Definition expr := state -> val.
Inductive HoaT(q:pname)(t:state):cmd -> asn -> asn -> Type:= 
| Asn: q;t$\triangleright$[Q[e / a]]a:=e[Q]
| Seq: q;t$\triangleright$[P]c1[Q] -> q;t$\triangleright$[Q]c2[R] -> q;t$\triangleright$[P]c1;c2[R] 
| If:  q;t$\triangleright$[P/\(isT b)]c1[Q] -> q;t$\triangleright$[P/\(isF b)]c2[Q] -> 
       q;t$\triangleright$[P] if b then c1 else c2 [Q]
| Cal: P$\Rightarrow$($\lambda$ s.(Pre p)s[!e/x!] /\ POrder(p,s[!e/x!])(q,t))-> 
       q;t$\triangleright$[P]a:=p(e)[...]
Equations interT{q:pname}{t:state}{c:cmd}{P Q:asn}
(H:q;t$\triangleright$[P]c[Q])(s:{v:state|P v}):{v:state|Q v/\v=s.1|Mod c}
by wf ((q,t), (size hoa)) (lexprod POrder lt) := 
interT (Asn e Q a) s:= $\square$ ((s.1)[e/a]);
interT (Seq c1 c2 P' Q' R' h1 h2) s:=
 let t:=(interT h1 s).1 in $\square$ ((interT h2 ($\square$ t)).1);
interT (If b c1 c2 P Q h1 h2) s =>   
 (match b(s.1) as r return b(s.1)=r->({v:state|Q v|...}) with
 |true =>fun cd=>$\square$ (interT h1 (exist _ (s.1)(conj s.$\textit{2}$ cd))).1
 |false=>fun cd=>$\square$ (interT h2 (exist _ (s.1)(conj s.$\textit{2}$ cd))).1
 end) eq_refl
interT (Cal q t p a e P imp1) s:=
 let ens:= (s.1)[!e/x!] in 
 let xs := (interT (PHoareT p ens) ($\square$ ens)).1 in
 ... .
Definition EQ (t:state) := fun s => s=t.
Parameter PHoareT:$\forall$ p t,p;t$\triangleright$[Pre p/\EQ t]Body p[Post p]
Inductive HoaP: cmd -> asn -> asn -> Type:= 
| Cast: (forall t, q;t$\triangleright$ [P/\EQ t]c[Q]) -> {P}c{Q}
Parameter PHoareP:$\forall$ p, {Pre p}Body p{Post p}
\end{lstlisting}





Here, we present a total-correctness logic for a simple language with recursive procedures but no global states. The logic---defined by the inductive type $HoaT$---is denoted by q;t$\triangleright$[P]c[Q]. The core novelty of this logic is to \textit{index} the local correctness derivation with its runtime entry, $(q,t)$, where $q: pname$ (the procedure name) and $t:state$ (the program state upon procedure entry), thus exposing the termination information in the \textit{surface}. We only discuss the call rule, $Cal$, since all the others are standard. We use $Pre~p$ and $Post~p$ for the pre- and post-conditions of the procedure $p$, respectively. The term $s[!e/x!]$ is supposed to be the entry state, constructed using the truncating substitution operator, i.e., $x$, the only parameter, is set to $e\ s$, and everything except for $x$ is set to a default value. $Cal$ requires the entry state to satisfy the precondition and decreases $POrder$, a well‑founded relation on entries. $POrder$ is supposed to model the runtime call graph; call-graph-based termination is typical in actual verifiers~\cite{leino2010dafny,Pandya1986Structure}, but we use it to define the definitional interpreter here. 
We omit discussion of the postcondition, which would be irrelevant to our technique.\footnote{Our Rocq development implements the adaptation~\cite{kleymann1999hoare} rule for local reasoning.}


%

The definition of the interpreter relies on the Equations plugin~\cite{sozeau2019equations}, which allows defining Rocq functions by well-founded recursion. Its signature matches the one in Section 1 (curly‑braced parameters are inferred automatically). The only addition is the second conjunct of the output, stating that the output state should be equal to the input state $s$ for all variables \textit{except} those modified by the command ($Mod\ c$). Such invariants are necessary to establish soundness. We define the interpreter by lexicographic order on (1) $POrder$, and (2) the structure of the derivation. In all cases except $Cal$, the entry stays the same but the structure decreases. In $Call$, the entry decreases.
We define the \textit{boxing} operator, $\square\ s$ (also written $\boxed{s}$), as a shorthand for the dependent pair constructor: we supply the state, let the predicate be inferred, and defer its proof as an obligation. $x.1$/$x.2$ accesses the first/second component of a pair. For each rule, we define the state‑transformer soundly; otherwise, Rocq reports an error immediately.

The case definitions are standard; we explain only \textit{If} and \textit{Cal}.

\textit{If}: We perform dependent pattern matching on $b(s.1)$ (b:  $state\rightarrow bool$), selecting a branch accordingly. The match evidence $cd$ is used to strengthen each branch’s precondition: we use $s.1$ as the state, and use the conjunction of $s.2$ and $cd$ as the proof. For each output state, we apply a boxing operation: although it satisfies $Q$, it modifies only $Mod\ c_1$ or $Mod\ c_2$, while $Mod\ c=Mod\ c_1\cup Mod\ c_2$. We argue that $Mod\ C$ over-approximates each case here. 


\textit{Cal}: Let $ens = s.1[!e/x!]$. We interpret the total‑correctness derivation for the procedure body ($PHoareT\ p$) at entry $(p, ens)$. 
Note the free fact $EQ~t$ in the precondition of $PHoareT$: the pre-state coincides with the entry state. This is essential to prove $POrder$ upon $Cal$ constructs. It should be clear that $POrder\ (p,ens)\ (q,t)$, establishing well-definedness. After the call, the final state is obtained via some state-transformer on the exit state. We omit it here. 

Finally, we mention the $Cast$ rule, it casts a family of total-correctness derivations into a partial-correctness one. This rule allows constructing partial-correctness derivation for methods ($PHoareP$) by casting $PHoareT$. The interpretation of the rule simply uses the total-correctness interpreter and is omitted here; see Section~\ref{sec:interpreter} for the interpretation of a more complicated cast rule. 

\subsection{Towards Real-world Hoare Logics}
\label{sec:towards}





\noindent \textbf{Mutable Heaps} 
Dynamic frames (DF) is a methodology for reasoning about mutable heaps. It has been studied via region logic and its variants~\cite{banerjee2008regional,banerjee2013local,bao2015conditional}, and implemented in verifiers such as Dafny~\cite{leino2010dafny,leino2009specification} and KeY~\cite{ahrendt2016deductive}. Unlike our earlier simple Hoare logic, which shallowly embeds assertions and expressions, region logic deeply embeds them. In this paper, we treat assertions as Boolean expressions for simplification. We explicitly define the semantics of expressions. We also define their \emph{footprints} to support local reasoning with heaps.

\noindent \textbf{Well-founded Functions} 
Because expression semantics and footprints are central in Hoare-logic reasoning, we want them to be total and unique: each expression has a single semantic value and footprint given a state. This is nontrivial, since expressions can call mutually recursive functions. We again rely on E-IDIs, reflecting the fact that practical verifiers use analogous termination arguments for functions and for terminating methods (functions must terminate, whereas methods need not).  Instead of total-correctness derivations for expressions, we define a metafunction to extract well-definedness constraints—the weakest precondition for an expression to be well-defined. This avoids a full deductive logic and is crucial for the \textit{well-definedness reflection} introduced later.

\noindent \textbf{Dynamic Dispatch}
Dynamic dispatch for both functions and methods is essential to object-oriented abstraction. This challenges our entry-indexing technique: when constructing total-correctness derivations or well-definedness proofs, only static entry information is available, yet we must ensure entry-indexed termination under dynamic dispatch. We address this by introducing virtual entries and behavioral subtyping between virtual and concrete entries.

\noindent \textbf{Total Correctness with Measures} In Section~\ref{sec:simple} we use assertions directly involving states (e.g., $POrder\ (p,s[!e/x!]) (q, t)$). However, in deep embeddings, assertions are interpreted over program values, which typically do not include states. We address this by formalizing \textit{measure-based termination}~\cite{leino2010dafny,ahrendt2016deductive}, which involves only relations on ordinary program values. 

 
\noindent \textbf{Outline} In the remainder of this paper, we develop partial- and total-correctness IDIs for a DF-based Hoare logic. In Section 3, we present the syntax and an example program. In Section 4, we define the semantics and footprints of expressions as two E-IDIs. In Section 5, we define a DF-based Hoare logic and its partial- and total-correctness IDIs, based on expression semantics and footprints. Most definitions of the expression semantics and footprints, as well as the DF-based Hoare logic, follow prior work~\cite{banerjee2008regional,banerjee2013local} closely. We focus on the novel challenges that arise in our development, as indicated in this section. For simplicity, we use standard mathematical notation; a mapping between the paper definitions and Rocq definitions is provided in the Rocq development.

\section{Language Definitions}


\subsection{Basic Definitions}
\label{sec:syntax}
\vspace{-0.2cm}
\begin{align*}
    & x,y,z\in Vars;\ f\in FieldNames;\ f,g,h\in FuncNames;\ m,n\in MethodNames\\ 
    & C,D\in\ ClassNames;\ T\in TraitNames;\ \\
    &Values: \ \ \ \ \ \ \ \ v\in \ N\ +\ B\ +\ Ptr\ +\ FSet\ Ptr\  \ \\
    &Exps:r,e,b,p,q::=\ x\ |\ v\ |\ e.f\ |\ ite\ b\ e_1\ e_2\ |\ominus\ e\ |\ e_1 \oplus\ e_2\ \\ 
    &\ \ \ \ \ \ \ \ \ \ \ \ \ \ \ \ \ \ \ \ \ \ \ \ \ \ |\ e\ is\ \ C\ |\ e\ is\ T\ |\ e_1.g@C(e_2)\ |\ e_1.g@T(e_2)\\
    &Commands: \ \ \ c ::=\ skip\ |\ x:= e\ |\ x.f:=e\ |\ x := C\ |\ if\ b\ then\ c_1\ else\ c_2  \\
    &\ \ \ \ \ \ \ \ \ \ \ \ \ \ \ \ \ \ \ \ \ \ \ \ \ \ \  |\ c_1; c_2\  |\ x:= y.m@C(\overline{z})\ |\ x:= y.m@T(\overline{z})\\
    &FuncSigs: \ \ \;\ \ \mathcal{G}::=fun\ g(x)\ requires\ p\ reads\ r\ decreases\ e\ ensures\ q\\
    &MethSigs: \ \;\ \mathcal{M}::=met\ m(\overline{x})\ requires\ p\ modifies\ r\ decreases\ (e|*)\ ensures\ q\\
    &Traits\ \  \ \ \ \,\ \ \ \ \ \ \mathcal{T}::=trait\ T\ \{\overline{\mathcal{G}};\ \overline{\mathcal{M}}\}\\
    &Classes: \ \ \ \ \ \ \; \ \mathcal{C}::=class\ C\ extends\ T\ \{\overline{f};\ \overline{\mathcal{G}\{e\}};\ \overline{\mathcal{M}\{c\}}\}
\end{align*}

Values consist of natural numbers, Booleans, pointers, and regions (i.e., finite sets of pointers). Pointers consist of pairs of the form $(address, class name)$, while $address$ is drawn from natural numbers. $\downarrow_n$, $\downarrow_b$, $\downarrow_p$, $\downarrow_r$ cast to a specified type, returning a default value if the input is not of that type. A state is a triple $(stack, heap, allocation\ set)$, where $stack$: $var \rightarrow val$, $heap$: $loc \rightarrow val$, and $allocation\ set$: $FSet\ Ptr$. Here, $loc$ are $(pointer, field name)$ pairs. The allocation set is supposed to record all allocated pointers. We use $alloc$ as a special variable that denotes the allocation set. Define $s[x]$: it returns the value of $x$ in the stack of $s$ if $x\neq alloc$, and the allocation set of $s$ otherwise. 

The expression forms are standard, except for two call constructs. Those correspond to the two possibilities of $e_1$'s static type (i.e., a class or a trait). 


The syntax of functions and methods follows Dafny's. But we model functions with only one parameter other than $this$, and fix the parameter name to be $x$. Functions without $x$ (i.e., only $this$) can be modeled by not using $x$ in their bodies. Preconditions, postconditions, read/write annotations, and measures are all expressions, though their intended types differ. Also note that the measure for a method may be \textit{*}, meaning that the method may not terminate. 



At last, we mention entries. An entry is made of an \emph{entry head}, which can be one of $C;m$, $T;m$, $C;f$, or $T;f$, and an entry state. The first two entry types are for methods, the last two are functions. Those with $T;m$ or $T;f$ as heads are \textit{virtual entries}, since traits do not contain implementations. However, they are important as placeholders for abstracting concrete entries, as introduced later. 


\subsection{A Running Example}
\label{sec:example}
\lstset{
  language=java,
  basicstyle=\ttfamily\small,
  morekeywords={class, trait, method, function, predicate, var, extends},
  keywordstyle=\color{blue}\bfseries,
  commentstyle=\color{gray}\itshape,
  stringstyle=\color{red},
  identifierstyle=\color{black},
  numbers=left,
  numberstyle=\tiny\color{gray},
  stepnumber=1,
  numbersep=5pt,
  backgroundcolor=\color{white},
  showspaces=false,
  showstringspaces=false,
  frame=single,
  tabsize=2,
  breaklines=true,
  emph={constructor, null, ensures, requires, reads, decreases, in, set, modifies, returns},
  emphstyle=\color{violet}\bfseries,
}

\begin{lstlisting}
trait Pizza{
    var fp: set <Pizza> // footprint
    function valid(): (ret:bool)
    decreases this.fp reads {this}+this.fp; 
    requires true // 4.1
    ensures ret==true==>this in this.fp // 4.2
    function price(): (ret:int)
    decreases this.fp reads {this}+this.fp; 
    requires valid() // 4.1
    method remA(f: set <Pizza>, p: int) returns (ret:Pizza) 
    decreases this.fp modifies this.fp// 5.1
    requires valid() && f==this.fp && p==this.price() // 5.1
    ensures ret.valid()&&(ret.fp<=f)&&ret.price()<=p} // 5.1
class Crust extends Pizza{
    function valid(): (ret:bool) {this in fp}
    function price(): (ret:int) {1}
    method remA(f: set <Pizza>, p: int) returns (ret:Pizza) 
    {return this;}}
class Anchovy extends Pizza{
    var nt: Pizza?
    function valid(): (ret:bool)
    {if this.nt == null then false else
    if this.nt in this.fp && this.nt is Pizza then 
    if this.nt.fp<this.fp && this !in this.nt.fp then
    this in this.fp && this.nt.valid() else false else false}
    function price(): (ret:int) {this.nt.price() + 1}
    method remA(f: set <Pizza>, p: int) returns (ret:Pizza)
    {var n:=this.nt; var f1:=n.fp; var p1:=n.price(); 
    var r:=n.remA(f1,p1); return r;}}
class Cheese extends Pizza{
    var nt: Pizza?  
    function valid(): (ret:bool){/*same as Anchovy.valid*/}
    function price(): (ret:int) {this.nt.price() + 1}
    method remA(f: set <Pizza>, p: int) returns (ret:Pizza)
    {var n:=this.nt; var f1:=n.fp; var p1:=n.price(); 
    var r:=n.remA(f1,p1); this.nt:=r; var f2:=r.fp; 
    this.fp:=f2 + {this}; return this;}}
\end{lstlisting}



This example is adapted from prior work~\cite{sun2024formalizing}. It models pizzas. \textit{Crust} is the base; \textit{Cheese} and \textit{Anchovy} can be added to an existing pizza. For simplicity, we omit the class annotations, which are the same as the trait's. The example is written in Dafny syntax; it differs slightly from Section~\ref{sec:syntax}, but the correspondence should be clear (we note only that $return\ e\triangleq ret:=e$). Following Dafny~\cite{leino2010dafny}, we split a function's preconditions into well-definedness and semantic preconditions, based on if they are necessary for the well-definedness of the function. All the function preconditions in the example are well-definedness preconditions. We first establish well-definedness (Section~\ref{sec:sem}), then use definitional axioms to prove semantic properties (Section~\ref{sec:definitional}).\footnote{In Dafny, semantic properties are established by first-order axioms. However, some semantic properties fall outside the solvers' scope and are still established internally. We do not have this concern, since we have higher-order definitional axioms.} 

The read/write \textit{annotations} are Dafny-style ``coarse-grained'' annotations: an entire region is marked as read or written. For example, the read annotation of $Pizza.valid$ means that all locations in $\{this\}+this.fp$ are possibly read.




The sole method, \textit{remA}, removes all Anchovies from a valid Pizza and returns the resulting Pizza. It takes two parameters (parameters are immutable), \textit{f} and \textit{p}, capturing the values of \textit{this.footprint} and \textit{this.price()} at entry. The postcondition requires that the result’s footprint be included in the input Pizza’s footprint and that its price not exceed the input’s price. We discuss these in Section~\ref{sec:hoare}.

\section{Expressions and Well-definedness}
In this section, we first define semantics and footprints as E-IDIs. Then, we indicate how semantic proofs can be done using the definitional axioms. 
\subsection{Semantics and Footprints}
\label{sec:sem}
To address the recursion problem, we introduce a well-definedness predicate and require the semantics and footprints to be total only for states that satisfy it. Formally, we define two metafunctions: $[\![e]\!]\ E: \{v:state|DF\ E\ e \ v=true\}\rightarrow val$ and $\{\![e]\!\}\ E: \{v:state|DF\ E\ e \ v=true\}\rightarrow FSet\ loc$, where $E$ is the entry of the evaluation. We emphasize that the footprint of an expression is a finite set of locations required to evaluate the expression. Here, $DF\ E\ e\ v$ checks that $e$ is well-defined in state $v$ given entry $E$. Note that $DF$ is decidable (returns $bool$, not $Prop$). This design allows \textit{well-definedness reflection}, introduced later. 



To define $DF$, we first set a well-founded order on entries, derived from the measures. Intuitively, measures reduce states into values (thus entries $EHead*state$ into reduced entries $EHead*val$), then a relation $F_v:(EHead*val\rightarrow EHead*val\rightarrow bool)$ determines the ordering on \textit{reduced entries}, where $EHead$ is the set of entry heads. Each program supplies $F_v$, while we define the meaning of measures to reduce states into values. However, a circularity arises: the semantics of the measures is not yet defined. We break this circularity via a simple but effective stratification: The semantics and footprint of \textit{call-free} expressions (denoted as $[\![\_]\!]_\alpha$ and $\{\![\_]\!\}_\alpha$) are defined using simple structural recursion (no entry-indexing). Measures need to be \textit{call-free}. With that, the relation on entries is defined as $FOrder (a\ b: entries)\triangleq F_v\ (a.1, [\![Mse\ a.1]\!]_\alpha\  a.2)\ (b.1, [\![Mse\ b.1]\!]_\alpha\ b.2)$, where \textit{Mse} returns the measure of an entry head. 



With $FOrder$, we can define $DF$. We present it for conditionals and calls.

$ite\ b\ e1\ e2$: When proving the $DF$ of either branch, the path condition $b$’s truth value is known. Here, circularity arises again, since $b$ is also an expression. We again require $b$ to be \textit{call-free}. With this, we can define $DF$ as a disjunction of the $DF$ condition of each branch, strengthened with the path condition. 
 
$e_1.g@T(e_2)$: We note that $e_1$ and $e_2$ must also be \textit{call-free}, since their semantics is important to define this case. With that, we first require the dynamic class extends the trait ($IsT\ C\ T$), the trait $T$ contains the function $g$ ($HasF\ T\ g$), and the function is \textit{kind\ 1} ($K1\ T\ g$, defined shortly). 
We also require the new entry $(T,g,s[!e_1/this;e_2/x!])$ be below the old entry $E$ in $FOrder$. However, we are not done yet. We must also require the well-definedness condition ($DFC\ T\ g$, which is again an expression) for this function. \textit{Call-freeness} is too restrictive for $DF$ conditions, since calling functions such as \textit{validity} are common for them. So we split functions into two kinds: (1) kind-1 (shape-regulating, e.g., validity), whose $DF$ conditions are \textit{call-free}; and (2) kind-2, whose $DF$ conditions may call kind-1 functions.
$DF$ requires all called functions to be \textit{kind-1}, so we use the \textit{call-free} semantic function to evaluate their well-definedness condition:
\begin{align*}
&DF\ E\ (ite\ b\ e_1\ e_2)\ s \triangleq \downarrow_b [\![b]\!]_\alpha\ s\ \&\&\ DF\ E\ e_1\ s\ ||\ \neg \downarrow_b\ [\![b]\!]_\alpha\ s\ \&\&\ DF\ E\ e_2\ s\\
&DF\ E\ (e_1.g@T(e_2))\ s \triangleq IsT(Cls ([\![e_1]\!]_\alpha\ s))\ T\ \&\&\  HasF\ T\ g\ \&\&\ K1\ T\ g\ \&\&\\
&let\ ens:=s[!e_1/this;e_2/x!]\ in\ FOrder\ (T,\ g,\ ens) \ E\ \&\& \downarrow_b [\![DFC\ T\ g]\!]_\alpha\ ens 
\end{align*}





Thus, we have two $DF$ ($DF$ and $DF_2$), two semantic metafunctions ($[\![\_]\!]_A$, and $[\![\_]\!]_\mathbf{A}$) and two footprint metafunctions ($\{\![\_]\!\}_A$, and $\{\![\_]\!\}_\mathbf{A}$). Nevertheless, those metafunctions are similar, only differing in call constructs. In particular, the definition of the call case in $DF_2$ uses the semantics $[\![\_]\!]_A$, defined via $DF$. We only present the \textit{kind-1} metafunctions here (for several important constructs). The reader is directed to the Rocq development for full definitions. 

We present the footprint definition for field access expressions, say, $e.f$, which is a \emph{singleton set} of the pair of the pointer value of $e$ and the field $f$:
\begin{align*}
&\{\![e.f]\!\}_A\ E\ s \triangleq \{(\downarrow_p([\![e]\!]_A\ E\ s), f)\}
\end{align*}


We present the semantics of calls, which is standard:
\begin{align*}
&\ \ \ \ \ \ \ \ [\![e_1.g@T(e_2)]\!]_A\ E\ s \triangleq let\ ens:= s.1[!e_1/this;e_2/x!]\ in\\ &
\ \ \ \ \ \ \ \ \ let\ C:=\ (Cls ([\![e_1]\!]_\alpha\ s.1))\ in\ [\![Body\ C\ g]\!]_A\ (C,g,ens)\ (\square\ ens)
\end{align*}
Here, we must prove $FOrder\ (C,g,ens)\ E$, but we only have $FOrder\ (T,g,ens)\ E$ in the $DF$ constraint. Thus, we require a proof obligation for each program:
\begin{align*}
&VirtualEntrySoundF:\ IsT\ C\ T\Rightarrow FOrder\ (T,f,s)\ E\Rightarrow FOrder\ (C,f,s)\ E 
\end{align*}
This holds trivially for the example program, since the measures are the same. 
To interpret $Body\ C\ g$, we must have $DF\ (C,g,ens)\ (Body\ C\ g)\ ens$, but we only have $\downarrow_b[\![DFC\ T\ g]\!]_\alpha\ ens$. Here, two obligations are required: 
\begin{align*}
&P1:K1\ T\ g\Rightarrow IsT\ C\ T\Rightarrow \downarrow_b [\![DFC\ T\ g\wedge this\ is\ C]\!]_\alpha\ s\Rightarrow\downarrow_b[\![DFC\ C\ g]\!]_\alpha\ s\\
&FDF1: K1\ T\ g\ \Rightarrow [\![DFC\ C\ g]\!]_\alpha\ s\Rightarrow DF\ (C,g,s)\ (Body\ C\ g)\ s
\end{align*}


To illustrate, we establish $FDF1$ for $Anchovy.valid$. The only non-trivial branch is the one with a recursive call ($this.nt.valid@Pizza()$). For that, we have the path condition $\downarrow_b[\![this.nt\neq null\wedge this.nt\ in\ this.fp\wedge this.nt\ is\ Pizza\wedge this.nt.fp<this.fp\wedge this\ !in\ this.nt.fp ]\!]_\alpha\ s$, we want to prove the proposition below, where we shorten $Pizza$ as $Pi$, and $Anchovy$ as $An$: 
\begin{align*}
&IsT(Cls ([\![this.nt]\!]_\alpha\ s))\ Pi\ \&\&\  HasF\ Pi\ valid\ \&\&\ K1\ Pi\ valid\ \&\&\ \\ 
&let\ ens:=s[!this.nt/this!]\ in\ FOrder\ (Pi,valid,ens)  \ (An,valid,s) \&\&  true
\end{align*}

We only explain the $FOrder$ conjunct. Here, the measure is $this.fp$, that conjunct reduces to $this.nt.fp<this.fp$, which is included in the path condition.

We introduce a few derived forms for the metafunctions:
\begin{enumerate}[topsep=0.1cm]
    \item Top-level: We omit the entry when it is $(Object, main,s)$, e.g., $[\![e]\!]_A\ (s,sok)\triangleq [\![e]\!]_A\ (Object,main,s)\ (s,sok)$ (and similarly for other metafunctions).
    \item Propositional: We cast the final results to $bool$, when expressions are used as propositions (e.g., as Hoare assertions). For $[\![e]\!]_A$, we further define $[\![e]\!]_B:state\rightarrow bool$, which returns true iff the expression’s well-definedness holds \textit{and} the casted Boolean result is true for \textit{any} well-definedness proof; we also define $[\![e]\!]_\mathbf{B}$ for $[\![e]\!]_\mathbf{A}$. That is, the expression \textit{reflects} its own well-definedness, providing a neat integration of well-definedness constraints and semantics. The definition relies on constructive UIP~\cite{hofmann1998groupoid}; see our Rocq development.
\end{enumerate}
\subsection{Proving with Definitional Axioms}
\label{sec:definitional}
Rocq (via Equations~\cite{sozeau2019equations}) automatically generates the definitional axioms for semantics and foorprints. We illustrate with the postcondition of $Pizza.valid$. 



We prove $[\![e_1.valid@Pizza()]\!]_\mathbf{B}\ s\Rightarrow [\![e_1\ in\ e_1.fp]\!]_\mathbf{B}\ s$. By the semantics of calls, we inspect $Body\ C\ valid$ with $C:=Cls\ ([\![e_1]\!]_\alpha\ s)$. Since we know $IsT\ (Cls\ ([\![e_1]\!]_\alpha\ s))\ Pizza$, we do a case analysis on the three classes implementing $Pizza$ and examine the function body. In each case, we can use the semantics of conditionals and conjunctions to show $[\![e_1\ in\ e_1.fp]\!]_\mathbf{B}\ s$. 

Some semantic proofs (e.g., the soundness of $read$ annotations) require induction. The Equations plugin offers functional induction principles; well-founded induction also applies because our metafunctions are defined by well-founded recursion. We use both as needed; see the Rocq development for details.

\section{Commands and Hoare Logic}
This section recalls syntactic effects, defines the executable region logic, and presents E-IDIs for total- and partial-correctness derivations.
\subsection{Syntactical Effects}
Effects are written $r\text{`}f$ with $r$ a call-free expression and $f$ a field name; $\eta$ and $\epsilon$ range over effects. In what follows, we primarily use lists of effects. The metafunction $locs:list\ effect\ \rightarrow state\rightarrow FSet\ loc$ computes the locations denoted by an effect list. We distinguish read and write effects: read effects over-approximate an expression’s footprint, while write effects over-approximate the locations a command writes. The metafunction 
$\mathit{reff}$ computes the read effects for an expression, while write effects are derived in Hoare logic (see Section~\ref{sec:hoare}).

As illustrated in Section~\ref{sec:example}, the $reads$ and $writes$ annotations concern regions. A region is converted to an effect list by pairing it with every field in the program (a finite set), using the metafunction $\mathcal{F}:expr\rightarrow list\ effect$.\footnote{This differs from Dafny, which uses universal quantification over fields.}

$\overline{\eta}\ \cdot/.\ \overline{\epsilon}$ denotes the separator predicate on states: it holds for a state $s$ if and only if the two effect lists denote disjoint sets of locations in $s$. The definition is standard~\cite{banerjee2008regional} and is omitted here. The notions of subeffects, immunity, and disjointness are also standard and should be clear from the definitions; we only note that the $!$ operator tests if two sets are disjoint.

\vspace{-0.5cm}
\begin{figure*}[h!]
\infax[subeffects]{\ P\Vdash \overline{\epsilon'} \leq \overline{\epsilon} \triangleq [\![P]\!]_\mathbf{B} \Rightarrow locs\ \overline{\epsilon'} \subseteq locs\ \overline{\epsilon} \ \ }
\infax[immune]{\overline{\epsilon_2}\ is\ P, \overline{\epsilon_1}\text{-}immune\triangleq \forall (r,f)\in \overline{\epsilon_2} .[\![P]\!]_\mathbf{B}\ \Rightarrow \mathit{reff}\ r \cdot/.\ \overline{\epsilon_1}}
\infax[disjoint]{\ \ \ \ \ \ \ \ \ \ \ \ \ \ \ \ \ \ \ \;  P\Vdash \overline{\epsilon}\ \# r'\triangleq \forall (r,f)\in \overline{\epsilon} .[\![P]\!]_\mathbf{B}\ \Rightarrow \downarrow_r [\![r]\!]_\alpha\ !\ \downarrow_r [\![r']\!]_\alpha}
\vspace{-1cm}
\end{figure*}
\subsection{Executable Region Logic}
\label{sec:hoare}




Figure 1 shows the rules. All but the last are stated in a total-correctness manner. Partial correctness follows by dropping the grey-background parts.
\begin{figure*}[t!]

\infrule[Assign]{x\notin fv\ e\ \cup fv\ P\ \ [\![P]\!]_\mathbf{B}\Rightarrow DF_2\ e}  {\cdarkg{D;n\triangleright}[P]x:=e[x=e\wedge P][]}  
\infrule[Write]{pure\ e\ \ call\text{-}free\ e}{\cdarkg{D;n\triangleright}[x\neq\ null]x.f:=e[x.f=e][[\{x\}\text{`}f]]} 
\infrule[Seq]{\cdarkg{D;n\triangleright}[P\wedge r=alloc] c_1 [Q][\overline{\epsilon_1}]\ \ \cdarkg{D;n\triangleright}[Q] c_2 [R][\overline{\epsilon_2}+\overline{\epsilon_2'}]\\ \overline{\epsilon_2}\ is\ P,\overline{\epsilon_1} \text{-}immune\ \ Q\Vdash \overline{\epsilon_2'} \# r\ \  \cdarkg{mse\notin Mod\ c_1}}{\cdarkg{D;n\triangleright}[P\wedge r=alloc] c_1;c_2 [R][\overline{\epsilon_1} + \overline{\epsilon_2}]}
\infrule[Frame]{\cdarkg{D;n\triangleright}[P] c [Q][\overline{\epsilon}]\ \ \overline{\eta}=\ \mathit{reff}\ R\ \ P\wedge R\Rightarrow \overline{\epsilon} \cdot /.\ \overline{\eta}\ \ Mod\ c\ !\ fv\ R}{\cdarkg{D;n\triangleright}[P\wedge R] c [Q\wedge R][\overline{\epsilon}]}
\infrule[Conseq]{\cdarkg{D;n\triangleright}[P'] c [Q'][\overline{\epsilon'}]\ \ [\![P]\!]_\mathbf{B}\Rightarrow [\![P']\!]_\mathbf{B}\ \ [\![Q']\!]_\mathbf{B}\Rightarrow [\![Q]\!]_\mathbf{B}\ \ P\Vdash \overline{\epsilon'}\leq \overline{\epsilon}}{\cdarkg{D;n\triangleright}[P] c [Q][\overline{\epsilon}]}
\infrule[CallT]{\overline{a}=Params\ T\ m \ \ |\overline{z}|=|\overline{a}| \ \ HasM\ T\ m\ \ [\![P]\!]_\mathbf{B}\Rightarrow [\![y\ is\ T\wedge Pre\ T\ m[y::\overline{z}/this::\overline{a}]\ ]\!]_\mathbf{B}\ \\  \cdarkg{Total\ T\ m}\ \ \cdarkg{[\![P]\!]_\mathbf{B}\Rightarrow \lambda s.\ M_v\ (T,m, [\![Mse\ T\ m]\!]_\alpha\  s[!y::\overline{z}/this::\overline{a}!])\ (D,n, s[mse])}}
{\cdarkg{D;n\triangleright}[P] x:=y.m@T(\overline{z})) [Post\ T\ m[x::y::\overline{z}/ret::this::\overline{a}]][\mathcal{F}(Wrt\ T\ m[y::\overline{z}/this::\overline{a}])]}

\infrule[Cast]{D;n\triangleright[P\wedge mse=Mse\ D\ n] c [Q][\overline{\epsilon}]\ \ mse\notin fv\ P\cup fv\ Q\cup fv\ \overline{\epsilon}}
{\{P\} c \{Q\}[\overline{\epsilon}]}
%
\caption{Rules of the Executable Region Logic}
\vspace{-0.5cm}
\end{figure*}

\noindent \textbf{Judgements}
Firstly, we note the meaning of the 5-tuple: $D;n\triangleright[P] c [Q][\overline{\epsilon}]$ means that using $D;n$ as entry head, $P$ as precondition, the execution of $c$ results in a state with $Q$ as postcondtion, and with write effects $\overline{\epsilon}$. Most of the rules follow region logic~\cite{banerjee2008regional,banerjee2013local}, to which we refer the reader for a detailed explanation. Here, we focus on explaining the new parts of our system. 

\noindent \textbf{$DF$ Constraints}
Only well-defined expressions have semantics and footprints, so we want any expression that occurs to be well-defined. This is ensured by rules like A\textsc{ssign}, W\textsc{rite}, where the precondition implies the well-definedness of the expression, or the expression is required to be \textit{call-free}. Pre- and post-conditions are also expressions. They are only used via $[\![\_]\!]_\mathbf{B}$, the \textit{$DF$ reflected semantics}. 

\noindent \textbf{Total Correctness and The Cast Rule}  First, those methods whose measures are \textit{not} \textit{*} are dubbed \textit{total methods}. Only total methods can be called in total-correctness derivations. 
We discussed measure-based termination for functions in Section~\ref{sec:sem}. Here, for methods, we define a well-founded order $MOrder$ following the definition of $FOrder$, and require $M_v$ similar to $F_v$. However, unlike $DF$, where entry states are carried everywhere and used for shallow well-founded constraints, here we must find a way to encode the well-foundedness constraint in deeply embedded assertions. We achieve this by assuming that at each entry, the measure would be evaluated and saved to an auxiliary variable $mse$. This eliminates the necessity of indexing derivations with entry states. With this, the termination condition at C\textsc{allT} can be phased as $M_v\ (T,m, [\![Mse\ T\ m]\!]_\alpha\  s[!y::\overline{z}/this::\overline{a}!])(D,n, s[mse])$, where $s[!y::\overline{z}/this::\overline{a}!]$ is the new entry state.


C\textsc{ast} expresses that a total-correctness derivation indexed by any entry head can be cast into a partial-correctness derivation. Here, we ensure that $mse$ is not used logically, since it would be modified to save the measure at the entry. 

\noindent \textbf{Method Verification} We require total correctness for total methods, and partial correctness for all methods: 
\begin{align*}
&MHoareT: Total\ C\ m\Rightarrow C;m\triangleright[P\wedge\ this\ is\ C\wedge mse=Mse\ C\ m] Bo[Q][\mathcal{F}\ \overline{\epsilon}]\ \ \ \ \ \ \ \ \ \ \ \ \ \\ 
&MHoareP: \{P\wedge\ this\ is\ C\} Bo[Q]\{\mathcal{F}\ \overline{\epsilon}\}\ \ \ \ \ \ \ \ \ \ \ \ \ \\ 
&where\ P=Pre\ C\ m, Q=Post\ C\ m, Bo=Body\ C\ m, \overline{\epsilon}=Wrt\ C\ m \ \ \ \ \ \ \ \ \ \ \ \ \ 
\end{align*}
Note that in addition to $Pre\ C\ m$, two free facts are used in the precondition: $this\ is\ C$ tells us the dynamic class of \textit{this}, allowing dynamically dispatching functions in assertions, as shown later. $mse=Mse\ C\ m$ asserts that $mse$ saves the measure at the entry, providing the necessary information for $mse$.   

Once $MHoareT$ is available, one could obtain $MHoareP$ for total methods using C\textsc{ast}, and only prove it manually for non-total methods. 

The following obligations ensure the soundness of virtual entries:
\begin{align*}
&TotalAbstraction: IsT\ C\ T\Rightarrow Total\ T\ m\Rightarrow Total\ C\ m\\
&VirtualEntrySoundM:\ IsT\ C\ T\Rightarrow MOrder\ (T,m,s)\ E\Rightarrow MOrder\ (C,m,s)\ E 
\end{align*}

\noindent \textbf{Example Derivation} Now, we illustrate $MHoareT$ for $Anchoy.remA$ in a forward reasoning style. We start with this precondition: 
$$
this.valid@Pi()\wedge f=this.fp\wedge p=this.price@Pi()\wedge\ this\ is\ An\wedge mse=this.fp
$$
Firstly, we unfold $this.valid@Pi()$ using the fact that $this\ is\ An$. After simplifying the conditionals, $this.valid@Pi()$ becomes: 
\begin{align*}
&this.nt\neq null\wedge this.nt\ in\ this.fp\ \wedge this.nt\ is\ Pi \wedge this.nt.fp< this.fp\wedge \ \ \ \ \ \ \\
&this\ !in\ this.nt.fp\wedge this\ in\ this.fp\wedge this.nt.valid@Pi()
\end{align*}
Similarly, by unfolding \textit{this.price@Pi()}, we get $p=this.nt.price@Pi()+1$. 

The first three statements are straightforward. We only note that the third statement requires a proof for $DF_2\ n.price@Pi()$,\footnote{The $DF$ conditions of the first two statements are $true$.} which is $n\ is\ Pi\ \wedge n.valid()@Pi$ after simplification. It holds since $n=this.nt$. The three statements add three equalities to the precondition: $n=this.nt\wedge f1=n.fp\wedge p1=n.price@Pi()$. 

For the fourth statement, we combine rules C\textsc{onseq} and F\textsc{rame} to simplify the conjuncts and split them into two groups. The first is the precondition for the call statement. The second group (show only a fragment) includes the conjuncts not used by the statement. The reader can check that (1) indeed satisfies the first implication in the premise of C\textsc{allT}. We only note the second implication, which essentially requires $n.fp< mse$ after simplification (recall that $this.fp$ is the measure of $Pi.remA$). (1) satisfies it trivially.   
\begin{align*}
& (1)\ n\ is\ Pi\wedge n.valid@Pi()\wedge f1=n.fp\wedge p1=n.price@Pi()\wedge n.fp< mse\\
&(2)\colive{f1< this.fp} \wedge \colive{f=this.fp}\wedge \ccyan{p=p1+1} \wedge this.nt\ is\ Pi \wedge this\ is\ An
\end{align*}

After the call, (1) is replaced with (1'), while (2) is preserved. 
$$
(1')\ r.valid@Pi()\wedge \colive{r.fp\leq f1}\wedge \ccyan{r.price@Pi()\leq p1} \ \ \ \ \ \ \ \ \ \ \ \ \ \ \ \ \ \ \ \ \ \ \ \ \ \ \ \ \ \ 
$$

This statement yields effects $[(n.fp)\text{`}fp;(n.fp)\text{`}nt]$, which are subeffects of $[(this.fp)\text{`}fp;(this.fp)\text{`}.nt]$ since $n=this.nt$ and $this.nt.fp< this.fp$. 

For the last statement, we add $ret=r$ to the conjunctions, then prove that the postcondition. $ret.valid@Pizza()$ follows from (1'), while $ret.fp\leq f$ follows from the \colive{olive} conjuncts. $ret.price()\leq p$ follows from the \ccyan{cyan} conjuncts. We also note that the write effects $[(this.fp)\text{`}fp;(this.fp)\text{`}.nt]$ are sound, since only the fourth statement yields write effects.  



\subsection{Interpreters for the Executable Region Logic}
\label{sec:interpreter}
Now, we show the interpretation of three core rules: C\textsc{all}T, F\textsc{rame}, and C\textsc{ast}. 


\noindent \textbf{Total-correctness Interpreter} As shown below, the signature of the total-correctness interpreter is similar to that in Section~\ref{sec:simple} but is significantly extended. First, note the second conjunct of the input state: we maintain the essential invariant that the auxiliary variable $mse$ always stores the measure at entry. Next, the second conjunct of the output state relates the output and input states. For such states, it requires that the two stacks are equal for all variables \textit{except} those modified by $c$ (excluding $alloc$); The two heaps are equal for all locations \textit{except} those denoted by $\overline{\epsilon}$; The two allocation sets are equal if \textit{alloc} is not modified, but if it is modified, the output set is a \textit{superset} of the input set.
\begin{align*}
&interT(H:D;n\triangleright[P]c[Q][\overline{\epsilon}])\ (s:\{v:state|[\![P]\!]_\mathbf{B}\ v\wedge (v[mse]=[\![Mse\ D\ n]\!]_\alpha\ t)\})\\
&:\{v:state|[\![Q]\!]_\mathbf{B}\ v\wedge (v=s.1|Mod\ c-\{alloc\};\overline{\epsilon};alloc\in Mod\ c)\}\\
&by\ wf ((D,n,t), size\ H) (lexprod\ MOrder\ lt)\\
&interT(CallT\ m\ x\ y\ \overline{z}\ T\ \overline{a}\ aeq\ i_1\ tot\ i_2)\ (s,\ sok) \triangleq\\
&\ let\ C:=Cls\ ([\![y]\!]_{\alpha}\ s)\ \ in\ let\ ens:=s[!y::\overline{z}/this::\overline{a}!][mse/Mse\ C\ m]\ in \ \\
&\ let\ xs := (interT\ (MHoareT\ C\ m\ \_)\ \boxed{ens}).1\ in\ \boxed{(s.1.1[xs[ret]/a], xs.1.2, xs.2)}\\
&interT(Frame\ c\ P\ Q\ R\ H_1\ \overline{\epsilon}\ \overline{\eta}\ dis_1\ dis_2)\ (s,\ sok)\triangleq\ \boxed{(interT\ H_1\ \boxed{s}).1}
\end{align*}
For C\textsc{all}T, the entry state is constructed first by using truncating substitution and then setting the $mse$ variable as the measure. With it, we interpret the method derivation.
To prove well-definedness, we have $M_v\ (T,m, [\![Mse\ T\ m]\!]_\alpha\  s[!y::\overline{z}/this::\overline{a}!]) (D,n, s[mse])$. We require that the measure do not depend on the variable $mse$. This gives $M_v\ (T,m, [\![Mse\ T\ m]\!]_\alpha\  ens) (D,n, s[mse])$. Then we use $VirtualEntrySoundM$ to get $M_v\ (C,m, [\![Mse\ T\ m]\!]_\alpha\  ens) (D,n, s[mse])$, as required. To use $MHoareT$, we must prove $Total\ C\ m$, which is guaranteed by $Total\ T\ m$ and $TotalAbstraction$. 
We must also ensure that $ens$ satisfies the precondition of $MHoareT$. This can be proved using the \textit{substitution lemma} and behavioral subtyping. For the final state, We use the heap and allocation set of the exit state, but the stack of the state before the call, assigned with $xs[ret]$. The proof for the postcondition and write effect also uses the substitution lemma, as well as behavioral subtyping. We omit it here. 

For the F\textsc{rame} rule, we interpret $H_1$ simply on the input state, i.e., $\boxed{s}$. The boxing operator there essentially drops the fact that $s$ also satisfies $R$, since $H_1$ requires only $P$. The output state is simply the output state of $H_1$, but we must re-establish the $R$ conjunct. Intuitively, The input and output state are equal except for $Mod\ c$ and $\overline{\epsilon}$. We know that the read effect list of $R$ is disjoint from the write effect list of $c$, and the set of free variables of $R$ is disjoint from $Mod\ c$. Thus, $R$ is preserved to the output state.


\noindent \textbf{Partial-correctness Interpreter} As shown below, the signature of the partial-correctness interpreter, $interP$, is similar to that of $interT$. Still, we add a step parameter and make the output options while removing the invariant on $mse$. 
\vspace{-0.1cm}
\begin{align*}
&interP (H:\{P\}c\{Q\}[\overline{\epsilon}]) (step:nat) (s:\{v:state|[\![P]\!]_\mathbf{B}\ v)\\
&:option\ \{v:state|[\![Q]\!]_\mathbf{B}\ v\wedge (v=s.1|Mod\ c-\{alloc\};\overline{\epsilon};alloc\in Mod\ c)\}\\
&interP\ (Cast\ D\ n\ H_T\ nin)\ n\ (s, sok) \triangleq\ let\ ins:=s[Mse\ D\ n/mse]\ in\ \ \ \ \ \ \ \ \ \ \ \\
&\ let\ outs\ :=\ (interT\ H_T\ \boxed{ins}).1\ in\ \boxed{outs[s[mse]/mse]}
\end{align*}
The semantics of C\textsc{ast} is simple: we set the variable $mse$ to the measure at the entry and then invoke the total-correctness interpreter. In the final state, we restore
$mse$ so that the set of modified variables remains equal to $Mod\ c$.


\section{Related Work}
\noindent \textbf{Definitional Interpreters}
Definitional interpreters were originally proposed to formalize the dynamic semantics of higher-order languages without recourse to domain theory, using only concrete functional languages~\cite{reynolds1972definitional,reynolds1998definitional}. Since then, they have been used for sequentializing languages~\cite{van2020principled}, dynamic languages~\cite{midtgaard2013engineering}, smart contracts~\cite{yang2019fether}, multi-level IRs~\cite{yu2025ratte}, and program analyses~\cite{darais2017abstracting}, among others.

For strongly typed languages, one must establish type safety by linking the typing relation to the interpreter. While this is often simpler with definitional interpreters than with other semantics techniques~\cite{amin2017type}, it still requires separate inductive proofs. The Intrinsic definitional interpreter methodology addresses this by writing interpreters that are directly guided by typing. IDIs have been used for various typing relations~\cite{bach2017intrinsically,rouvoet2020intrinsically,van2022intrinsically}. Our work is the first to apply the IDI methodology to Hoare logics rather than to typing relations.

\noindent\textbf{Dynamic Frames and Dynamic-frames-based Hoare Logics}
Dynamic Frames (DF) is a general methodology for reasoning about imperative programs. Introduced in a refinement setting~\cite{kassios2006dynamic,kassios2011dynamic}, it has been widely used in Hoare logics and in practical verifiers~\cite{leino2010dafny,ahrendt2016deductive}. The most notable DF-based Hoare logics are region logic and its variants~\cite{bao2015conditional,bao2018unifying,banerjee2013local,banerjee2013local2,banerjee2023relational,banerjee2018logical}. Our logic, XRL, follows the region logic formalism but extends it with well-founded functions, behavioral subtyping, and total correctness. The work by Banerjee et al.~\cite{banerjee2018logical} is the only work we know that formalizes well-founded functions and proves soundness. But their development is parameterized on a \textit{candidate interpretation}, while our work can be viewed as a concrete instantiation of it. Furthermore, XRL is based on first-order Hoare-logic rules to support method calls, whereas prior region logics (and most Hoare logics with recursive methods) rely on second-order rules~\cite{banerjee2013local2,banerjee2018logical}.

DF-based Hoare logics and verifiers differ in syntactic effects. In this paper, we follow Dafny in using coarse-grained effects~\cite{leino2010dafny}. However, our footprint is defined semantically, and the key lemmas are proved at that level; they do not depend on any particular syntactic effects. Therefore, we believe that alternative syntactic effects can also be readily layered on top of the semantic footprint.

The circularity problem is a long-standing issue in Hoare logics with functions~\cite{Darvas2007Practical,leino2009proving}. Our stratification is inspired by Dafny~\cite{leino2010dafny}, but it fixes three levels, whereas Dafny permits arbitrarily many.

\section{Conclusion and Future Work}
In this work, we present definitional interpreters for Hoare logic, yielding the first DF-based Hoare logic with well-founded functions, behavioral subtyping, and total correctness, as well as the first fully mechanized DF-based Hoare logic.

An interesting future work is to link our definitional axioms for expressions with Dafny’s axioms to argue their soundness, and to use our axioms to extend Dafny's axioms. The second would support more automatic reasoning—class-branching over implementing classes and well-founded/functional induction—both are missing in Dafny today. Other future work includes supporting more PL features (e.g., inheritance) and studying IDI’s applicability to separation logic.






\bibliography{samplebib}
\bibliographystyle{splncs04}
\end{document}